\documentclass{ws-brl_2}
\usepackage[super,sort,compress]{cite}
\usepackage{xcolor}
\usepackage[verbose]{hyperref}
\usepackage{breqn}
\hypersetup{colorlinks=false,allbordercolors=blue,pdfborderstyle={/S/U/W 1}}

\begin{document}

\markboth{Yusuke K. Shibasaki}
{A Loewner-Theoretic Approach to the Nonlinear Generalized Langevin Equation}

\catchline{}{}{}{}{}

\title{A Loewner-Theoretic Approach to the Nonlinear Generalized Langevin Equation: The Role of Entropy in Colored Noise Environment}

\author{Yusuke K. Shibasaki}

\address{College of Art, Nihon University, Nerima, Tokyo, 176-8525 Japan 
\email{shibasaki.yusuke@nihon-u.ac.jp}}

\maketitle

\begin{history}
\received{Day Month Year}
\revised{Day Month Year}
\accepted{Day Month Year}
\published{Day Month Year}
\end{history}

\begin{abstract}
In this study, the formal derivation of a one-dimensional nonlinear generalized Langevin equation is demonstrated using a decomposition method based on the conformal transformation governed by the chordal Loewner equation. Here, we used a modified Mori-Zwanzig method whose operator is substituted by that is derived from the discrete Loewner evolution. By this approach, the different types of fluctuation-dissipation relation (FDR) were reformulated using mathematical terms affected by the conformal maps. Dealing with a memory kernel that models the cell migration experiment, the numerical simulation was performed to obtain the specific scaling law of energy dissipation that is common among the two obtained types of FDRs. In addition, the concept of Loewner entropy is used for the estimation of the canonical ensemble in the colored noise environment throughout the theoretical analyses. 

\end{abstract}

\section{Introduction}	
As for the recent theoretical developments of the statistical-mechanical theory for the Brownian motion, diffusion processes, and the random-walks appearing in the experimental settings, the stochastic characteristics of the dissipative structures of the physical systems have been gradually clarified \cite{limmer2024, mézard2009, boffetta2026}. The Langevin equation is one of the fundamental equations of the non-equilibrium statistical physics, which has been studied in the context of the Einstein’s theory of Brownian motion \cite{nelson1967}, closely related to the Fokker-Planck equation \cite{limmer2024, mézard2009, boffetta2026, risken1989}. The original model of the Langevin equation is driven by a stochastic term that is expressed by the white noise while the celebrated Kubo’s formula yields a fluctuation-dissipation relation (FDR) calculated from the autocorrelation function of the dynamical variable that we observe \cite{kubo1957, marconi2008, sarracino2019, risken1989}. The Brownian motion in a non-equilibrium state often exhibits anomalous behaviors that include some memory effects in its dynamics; this type of the anomalous diffusion model has been applied to physical, chemical and biological systems in terms of the generalized Langevin equation (GLE) \cite{risken1989, mori1965, zwanzig1973, vrugt2020, chung2019, cairano2022, hu1990, kaulakys2004, mitterwallner2020, netz2024, klimek2025, vroylandt2022, hadeler2004, mckinley2018, cui2018, huang2008, Taloni2014}. Because the GLE is implemented to describe the non-equilibrium phenomena, its statistical-physical theory intrinsically includes the essential problems concerning the reduction of nonlinearity\cite{risken1989, mori1965, zwanzig1973}. In this context, the Mori-Zwanzig formalism based on the projection operator method is one of the most successful theoretical results in deriving and analyzing the nonlinear GLE \cite{mori1965, zwanzig1973}. In this method, by introducing a specific operator associated with the Langevin equation, the decomposition of the memory effect term and noise term is performed \cite{mori1965, zwanzig1973}. This enables a physical interpretation of the GLE in a reliable manner; however, the prediction of the behavior of the systems remains to be a difficult issue because the dynamics of the main variables are highly nonlinear. Current studies suggest that the GLE is widely used model for diffusion processes in physics, chemistry and biology, while the notable real-world applications include Molecular Dynamics (MD) simulations \cite{huang2008}, cell migration modeling \cite{mitterwallner2020, klimek2025}, chemical network modelling., etc. \ \
 
In this study, we shall demonstrate a conformal mapping-based approach to the theoretical treatment of the nonlinear GLE using the Loewner differential equation (LDE) \cite{lawler2008, rohde2005}. As mentioned in the previous studies, the statistical-physical approach using LDE is based on the mixing dynamical system’s property of the iterated conformal maps \cite{shibasaki2020, shibasaki2025}. By an analogy with Mori-Zwanzig method, we observe that this property is also used for the formalism of the GLE, while providing a novel viewpoint for the approaches to non-equilibrium statistical physics. Particularly, using this Loewner-theoretic approach, we shall discuss the comparison of the different types of FDRs (i.e., one is Kubo type one and the other is the one directly obtained from Loewner theory). The numerical simulation is also performed to observe the scaling law in the energy dissipation by applying the memory kernel that models the cell migration dynamics. This paper consists of the following parts. In Sec. 2, the author introduces a modified Mori-Zwanzig formalism using the transfer operator of the conformal maps of the Loewner equation. Subsequently, in Sec. 3, we compare three different types of FDR for the generalized Langevin equation and discuss its applicability to the phenomena in physical and biological systems. In Sec.4, we perform numerical simulations to test the present theoretical results. In Sec. 4 and Sec. 5, the discussion and conclusion drawn from this study are remarked.

\section{Modified Mori-Zwanzig formalism}
In this section, we shall introduce model equations that we analyse in this paper. First, we consider the chordal Loewner equation, which is used for operator analyses in the main result. Let us consider the 2-dimensional (2D) growth of a simple curve $\gamma_{[0,s]}$ on the upper half complex plane $\mathbb{H}$. The chordal Loewner equation is described by: 
\begin{equation}
\frac{\partial g_s(z)}{\partial s}=\frac{2}{g_s\left(z\right)-U_s},\ \ \ \ \ \  g_0\left(z\right)=z\in\mathbb{H}.\ \ \ \ 
\end{equation}
Here, $g_s$ is a conformal map which transforms the region $\mathbb{H}\setminus\gamma_{[0,s]} $to $\mathbb{H}$. The term $U_s$ is a one-dimensional real-valued function called the Loewner driving function. The 2D curve $\gamma_{[0,s]}$ and the function $U_s$ are parametrized by the capacity $s$ which works as a specific time parameter describing the curve growth in $\mathbb{H}$. In a discretized expression \cite{shibasaki2020, shibasaki2025, kennedy2008}, it is known that the relation between the curve $\gamma_{\left[0,s\right]}:=\{z_0\left(=0\right),\ z_1,z_2,\ldots,z_n,\ \ldots,\ z_N\}$ and driving function $U_s$ is expressed as the transformation of the variable $w_n(=\Delta U_n+2i\sqrt{\Delta s_n})$:
\begin{equation}
\begin{split}
w_1&=z_1, w_2=h_1\left(z_2\right), …,  \\  
w_n&=h_{n-1}\circ h_{n-2}\circ\cdots\circ h_1\left(z_n\right), …, \\ 
w_N&=h_{N-1}\circ h_{N-2}\circ\cdots\circ h_1\left(z_N\right).
\end{split}
\end{equation} 
where 
\begin{equation}
h_n\left(z\right)=\sqrt{\left(z-\Delta U_n\right)^2+4\Delta s_n}.\ \ \ 
\end{equation} 
Let us define the probability density function of $\eta_s\left(n\right)=\Delta U_n/\Delta s_n$ as $f\left(z\right)\in \mathbb{C}$. We introduce a transfer operator $\mathcal{P}_n$ defined as \cite{shibasaki2025, beck1995, ruelle2004}:
\begin{equation}
\mathcal{P}_nf\left(z\right)=\sum_{z\in h^{-1}(w)}{\left|{h^\prime}_n\left(\chi_\sigma\left(w\right)\right)\right|^{-1}f(\chi_\sigma\left(w\right))},\  
\end{equation} 
where the prime of $h_n$ denotes the spatial derivative and
\begin{equation}
\chi_\sigma\left(w\right)= h_{\sigma=1,2}^{-1}=\Delta U_n\pm\sqrt{w^2-4\Delta s_n}
\end{equation} 
is preimage of $h_n\left(z\right)$. 

Subsequently, we discuss the derivation of the nonlinear GLE from the energy of the system. Let us denote the position and velocity of a particle at time $t$ as $x_n(t)$ and $v_n(t)$ and assume that the system has a Hamiltonian function $H(x)$. We consider that the evolution of a function $u\left(t,x\right)$ associated with this system is represented by an operator $\mathcal{L}$, that is: \cite{chung2019}
\begin{equation}
\frac{\partial}{\partial t}u\left(t,x\right)=\mathcal{L}u\left(t,x\right).\\ 
\end{equation}
Using the relation in Eq. (6), we obtain the following. 
\begin{equation}
u\left(t,x\right)=\left(e^{t\mathcal{L}}q\right)\left(x\right).
\end{equation}
Substituting Eq. (7) into Eq. (6), it follows that
\begin{equation}
\frac{\partial}{\partial t}\exp{\left(t\mathcal{L}\right)}q\left(x_n,v_n,t\right)=\mathcal{L}\exp{\left(t\mathcal{L}\right)}q\left(x_n,v_n,t\right)=\exp{\left(t\mathcal{L}\right)}\mathcal{L}q\left(x_n,v_n,t\right).
\end{equation}
Here, $q$ is an arbitrary function. For the one-dimensional dynamics, we should also notice that:
\begin{equation}
\frac{\partial}{\partial t}\exp{\left(t\mathcal{L}\right)}x_n=\exp{\left(t\mathcal{L}\right)}\mathcal{L}x_n=\exp{\left(t\mathcal{L}\right)}\mathcal{L}x_n
\end{equation}
We define a new operator as:
\begin{equation}
\mathcal{Q}_n:=1-\mathcal{P}_n
\end{equation}
From Dyson’s operator identity \cite{chung2019, vrugt2020} , we obtain
\begin{equation}
\frac{\partial}{\partial t}\exp{\left(t\mathcal{L}\right)}x_n=\exp{\left(t\mathcal{L}\right)}\mathcal{L}x_n+\int_0^t \exp[(t-t')\mathcal{L}]\mathcal{P}_n \mathcal{L}\exp(t\mathcal{Q}_n\mathcal{L})\mathcal{Q}_n\mathcal{L}x_ndt^\prime\ 
\end{equation}
Consequently, we obtain the following time-differential equation \cite{chung2019}: 
\begin{equation}
\frac{\partial}{\partial t}x_n\left(t\right)=h_n\left(x\right)+\int_{0}^{t}{K_n\left(t^\prime, x\right)dt^\prime}+F_n\left(t,x\right).\ \ \ \ 
\end{equation}
where
\begin{equation}
F_n\left(t,x\right)=\exp{\left(t\mathcal{Q}_n\mathcal{L}\right)\mathcal{Q}_n\mathcal{L}x_n}
\end{equation}
and 
\begin{equation}
K_n\left(t^\prime, x\right)=\mathcal{P}_n\mathcal{L}F_n\left(t,x\right).\ \ \  
\end{equation}
The equations (12)-(14) are read as a nonlinear GLE because the term $F_n\left(t,x\right)$ and $K_n\left(t^\prime,\ x\right)$ are noise and memory effect terms. More explicitly, we obtain the following expression. 
\begin{equation}
\frac{\partial}{\partial t}x_n\left(t\right)=h_n\left(x\right)+\int_{0}^{t}{\eta_s^M(t^\prime)dt^\prime}+F_n\left(t,x\right).\ \ \ \ \ \ \ \ 
\end{equation}
Here, $\eta_s^M(t^\prime)$ is the Loewner driving force of the curve associated with $\mathcal{L}F_n\left(t,x\right)$, and the second term of the right-hand side represents the memory effect in the dynamics.
\begin{equation}
\frac{\partial}{\partial t}v_n\left(t\right)=h_n\left(x\right)+\int_{0}^{t}{\eta_s^M(t^\prime)v_n\left(t\right)dt^\prime}+F_n\left(t,x\right).\ \ \ \ \  
\end{equation}
The term $F_n\left(t,x\right)$ corresponds to the correlated noise term, that is, the colored noise effect. From the fluctuation-dissipation theorem in Einstein’s sense, we obtain the following restriction \cite{kubo1957, cairano2022}:
\begin{equation}
K_n\left(t^\prime, x\right)=\eta_s^M\left(t^\prime\right)=\left\langle F_n\left(t,x\right) F_n\left(t^\prime,x\right)\right\rangle.\ \ \ 
\end{equation}
Here, the brackets denote the ensemble average. For the above, the derivation of the GLE of a one-dimensional dynamic $x_n\left(t\right)$ from an energy function (i.e., Hamiltonian) is demonstrated. As we showed, in this method, we use a conformal transformation governed by the Loewner differential equation (LDE) instead of the conventional projection operator. For this approach, the mapping included in the operator $\mathcal{P}_n$ is obtained from conformal mapping and this enables the calculation of the memory term $K_n\left(t^\prime,\ x\right)$ from colored noise term $F_n\left(t,x\right)$. Therefore, the forms of the decompositions in Eqs. (15) and (16) are different from those of the Mori-Zwanzig method; however, the present approach becomes a unique method when we consider the restriction of FDR in Eq. (17). In addition, the statistical-mechanical theory for the GLE also is reformulated because the present method relies on the conformal transformation in the complex plane. In the next section, we shall derive the fluctuation-dissipation relations (FDRs) in Kubo’s sense and a generalized FDR based on the dynamical system theory.
 
\section{Comparison of Three Types of Fluctuation-Dissipation Relations}
In this section, we compare two different types of fluctuation-dissipation relations (FDRs) for the generalized Langevin equation (GLE) that is derived from a Hamiltonian in Sec. 2. First, we show that the straightforward calculation from the derived GLE in the previous section leads to a response function formalism in a conventional form. We consider the autocorrelation function of the variable $v_n(t)$ as follows:
\begin{dmath}
\left\langle v_n\left(t\right)v_n\left(t+h\right)\right\rangle_{eq}  = \left\langle\left(h_n\left(x,t\right)+\int_{0}^{t}{\eta_s^M\left(t^\prime\right)dt^\prime}  
+  F_n\left(t,x\right)\right)\left(h_n\left(x,t+h\right) +\int_{0}^{t+h}{\eta_s^M\left(t^\prime\right)dt^\prime}+F_n\left(t+h,x\right)\right)\right\rangle_{eq}
\end{dmath}
For convenience, we define a random walk $V_s$ as:
\begin{equation}
V_s=\int_{0}^{t+h}{\eta_s^M\left(t^\prime\right)dt^\prime}.
\end{equation} 
The autocorrelation function is calculated as: 

\begin{dmath}
\left\langle v_n\left(t\right)v_n\left(t+h\right)\right\rangle_{eq}\simeq\left\langle\eta_s\left(t\right)\eta_s\left(t+h\right)\right\rangle_{eq}=\left\langle h_n\left(x,t\right)h_n\left(x,t+h\right)\right\rangle_{eq}+\left\langle h_n\left(x,t\right)V_s\left(t+h\right)\right\rangle_{eq} +\left\langle h_n\left(x,t+h\right)V_s\left(t\right)\right\rangle_{eq}+\left\langle F_n\left(t\right)F_n\left(t+h\right)\right\rangle_{eq}\
\end{dmath} 
We here used the approximation of $v_n\left(t\right)\simeq\eta_s\left(t\right)$ from the property of the dynamics of the backward Loewner evolution (See, Appendix. B). We also note that the ensemble average is taken over the equilibrium distribution of the Loewner driving force. Thus, it immediately follows that the response function $R\left(t, t+h\right)$ of FDR is led by Kubo’s formula as \cite{shibasaki2022, shibasaki2024}:  
\begin{dmath}
R\left(t, t+h\right)  \\
=\alpha d(x,t)\left\{\left\langle h_n\left(x,t\right)h_n\left(x,t+h\right)\right\rangle_{eq}+\left\langle h_n\left(x,t\right){\frac{\partial}{\partial s}}V_s(t+h)\right\rangle_{eq}+\left\langle h_n\left(x,t+h\right)\frac{\partial}{\partial s}V_s(t)\right\rangle_{eq}+\left\langle F_n(t)F_n(t+h)\right\rangle_{eq}\right\} 
=\alpha d\left(x,t\right)\left\{\left\langle h_n\left(x,t\right)h_n\left(x,t+h\right)\right\rangle_{eq}+\left\langle h_n\left(x,t\right)\eta_s^M\left(t+h\right)\right\rangle_{eq}+\left\langle h_n\left(x,t+h\right)\eta_s^M\left(t\right)\right\rangle_{eq}+\left\langle \eta_s^M\left(t+h\right)\right\rangle_{eq}\right\}
\end{dmath}
where 
\begin{equation}
d\left(x,t\right)=\frac{2t^\prime}{{v_n\left(t^\prime\right)}^2+{t^\prime}^2}.
\end{equation}
and $\alpha$ is a suitable constant. In the second line of Eq. (21), we used the relation in Eq. (17). The above result is another expression of the FDR for the system in Eq. (16), and it describes the nonlinear response of the system in a more detailed manner. For the above discussion, we also notice that the energy dissipation of the system obeys the following scaling relation: 
\begin{equation}
R\left(t,\ t+h\right) \sim C\left(h\right)t^{-1},\\
\end{equation}
where $C(h)$ is a suitable constant depending on $h$. This scaling law corresponds to a type of universal class found in the self-organization systems \cite{hu1990, kaulakys2004}. 

For the further investigation, we define the entropy of probability distribution function of the Loewner driving force $\eta_s(t)$ associated with $v_n(t)$ expressed as \cite{shibasaki2025, shibasaki2024}:
\begin{equation}
S_{Loew}=-\ln{p\left(\eta_s\left(t\right)\right)}.\ \ 
\end{equation}
Here, the probability distribution of driving force $p(\eta_s\left(t\right))$ is analogous to the microcanonical ensemble of the dynamics of the variable $v_n(t)$. Using this fact, the response function $R\left(t,\ t+h\right)$ in Eq. (21) is rewritten as:
\begin{dmath}
R\left(t,\ t+h\right)=\alpha d\left(x,t\right)\sum_{n=0}^{N}\left\{h_n\left(x,t\right)h_n\left(x,t+h\right)+h_n\left(x,t\right)\eta_s^M\left(t+h\right)+h_n\left(x,t+h\right)\eta_s^M\left(t\right)+F_n\left(t\right)F_n\left(t+h\right)\right\}\exp{\left(-S_{Loew}\right)}.\ \ \ \ \ \ \ \ \ \  
\end{dmath}
This is also the nonlinear response function derived in accordance with Kubo’s formula. 

Contrary to the above approach, the previous studies \cite{shibasaki2022, shibasaki2024} have suggested a method of deriving the FDR directly from the one-dimensional dynamics using the conformal time transformation. According to this method, the time parameter of the Langevin equation is converted as $t\rightarrow s$. For this change, the dynamics of $v_n\left(s\right)$ is rewritten as \cite{shibasaki2022, shibasaki2024}:
\begin{dmath}
\frac{\partial}{\partial s}v_n\left(s\right)=\left(\frac{2t^\prime}{{v_n\left(t^\prime\right)}^2+{t^\prime}^2}\right)h_n\left(x\right)+\left(\frac{2t^\prime}{{v_n\left(t^\prime\right)}^2+{t\prime}^2}\right)\int_{0}^{t}{\eta_s^M(t^\prime)dt^\prime}+\left(\frac{2t^\prime}{{v_n\left(t^\prime\right)}^2+{t^\prime}^2}\right)F_n\left(t,x\right). 
\end{dmath} 
Because the terms in the right-hand side tend to be zero as $s$ becomes large, we obtain the equilibrium distribution.
\begin{equation}
p_{eq}\left(v_n\left(s\right)\right)=\frac{1}{Z}\exp{\left(-\ln{S\left(v_n,t\right)}\right)}\  
\end{equation} 
where
\begin{equation}
Z=\sum_{n=0}^{N}{\exp(-\ln{S\left(v_n,t\right)})}.\ \ \
\end{equation}
Here, $S\left(v_n,t\right)$ is a non-negative function. The dynamics of $v_n$ is converted into linear ones. Therefore, the nonlinear response function derived from the direct Loewner-theoretic approach is expressed as:
\begin{equation}
R\left(s,\ s+h\right)=\alpha^\prime d\left(x,t\right)\left\langle v_n(s)\frac{\partial}{\partial s}v_n(s+h)\right\rangle_{eq}.\ 
\end{equation}
Here, $\alpha^\prime$ demotes a suitable constant. The ensemble average of the above equation is rewritten as: \cite{shibasaki2025}
\begin{equation}
R\left(s,\ s+h\right)=\alpha^\prime d\left(x,t\right)\sum_{n=0}^Nv_n(s)\Delta v_n(s+h)\exp{\left(-S_{Loew}\right)}.     
\end{equation}
The above expression also reads as:
\begin{equation}
R\left(t,\ t+h\right) \sim C^\prime\left(h\right)t^{-1}.\ 
\end{equation}
The above direct approach is useful especially when we do not have enough knowledge about the mechanism behind the observed stochastic dynamics; however, the FDR derived from this approach assumes the nonlinear response of the system is forced in a conformal field. Therefore, the comparison between three types of response function that we derived is important to examine the validity of the theory in real physical behaviors.

\section{Numerical Simulation: Cell Migration as a Biophysical Example }
As an example of the real-world behavior of the generalized Langevin dynamics, we here consider the cell migration dynamics. In accordance with the experimental results \cite{mitterwallner2020, klimek2025} , the kernel function of cell migration dynamics is assumed to be:
\begin{equation}
K_n\left(t^\prime,\ x\right)= b+\exp{\left(-\frac{t}{c}\right)}.\\ 
\end{equation}
We should also remind that\\ 
\begin{equation}
F_n\left(t,x\right) \sim \exp{\left(-\frac{t}{2c}\right)}
\end{equation}
so that the memory kernel in Eq. (32) satisfies the condition in Eq. (17). In the discrete time scheme, Eq. (16) is approximated by the following equations:
\begin{dmath}
v_n(j)=v_{n}(j-1)+\tau h_n\left(x_n(j-1)\right)+ \sum_{j=0}^N [b+\exp(-t/c)]\sqrt\tau W_{n}(j-1)v_{n}(j-1)+\sqrt{\tau [b+\exp(-t/2c)]}W_{n}(j-1), 
\end{dmath}
where $x_n$ is calculated as  
\begin{dmath}
x_n= x_{n}(j-1)+\tau v_{n}(j-1), \\
\end{dmath}
and we chose $h_n$ as 
\begin{dmath}
h_n\left(x_n\right)=-\frac{1}{2}k{x_n}^2
\end{dmath}
where $k$ is a constant.

The numerical simulation was performed to confirm the bahaviors of the nonlinear response function by Kubo-type formula in Eqs. (21) and that by direct Loewner-theoretic approach in Eq. (29). In order to observe the FDRs for, we first computed the functions $r(t, t+h)$ and $r(s, s+h)$, each of which represents the responses of the individual trajectories. To calculate the ensemble averages, 100 realizations of $r(t, t+h)$ and $r(s, s+h)$ are generated. It means that the calculation of $R(t, t+h)$ and $R(s, s+h)$ were performed by taking the ensemble averages of 100 realizations of $r(t, t+h)$ and $r(s, s+h)$, respectively. We compute the term $\eta_s^M$ by the zipper algorithm using the vertical slit map \cite{kennedy2008}, and this term is used in the calculation of $R(t,t+h)$ in Eq. (21). Figure 1(a) shows the time-dependent behavior of $R(t, t+h)$. As the theory suggested, the decay of $R(t, t+h)$ was observed and it suggests the energy dissipation of the the dynamics of GLE. The similar behavior was found in the plot of $R(s, s+h)$, which is shown in the figure 1(b). The log-log plots suggest that the decay is scaled as $t^{-1}$ in both $R(t, t+h)$ and $R(s, s+h)$. Thus, the numerical results are consistent with Eqs. (23) and (31). The parameter used for these results of the simulation is $b = 0.5$, $c = 2.0$, $k = 1.0$, $\tau = 0.00001$, where $t=j\tau$. The function $R(t, t+h)$ is calculated by fixing $t=1\tau$ and $s=1\tau$.  The stationary property of the computed $\eta_s^M$ was also confirmed, and it reflects the important property of the operator $\mathcal{P}_n$ used in the modified Mori-Zwanzig method we described in Sec. 3.  

\begin{figure}[htbp]
\centerline{\includegraphics[width=13cm]{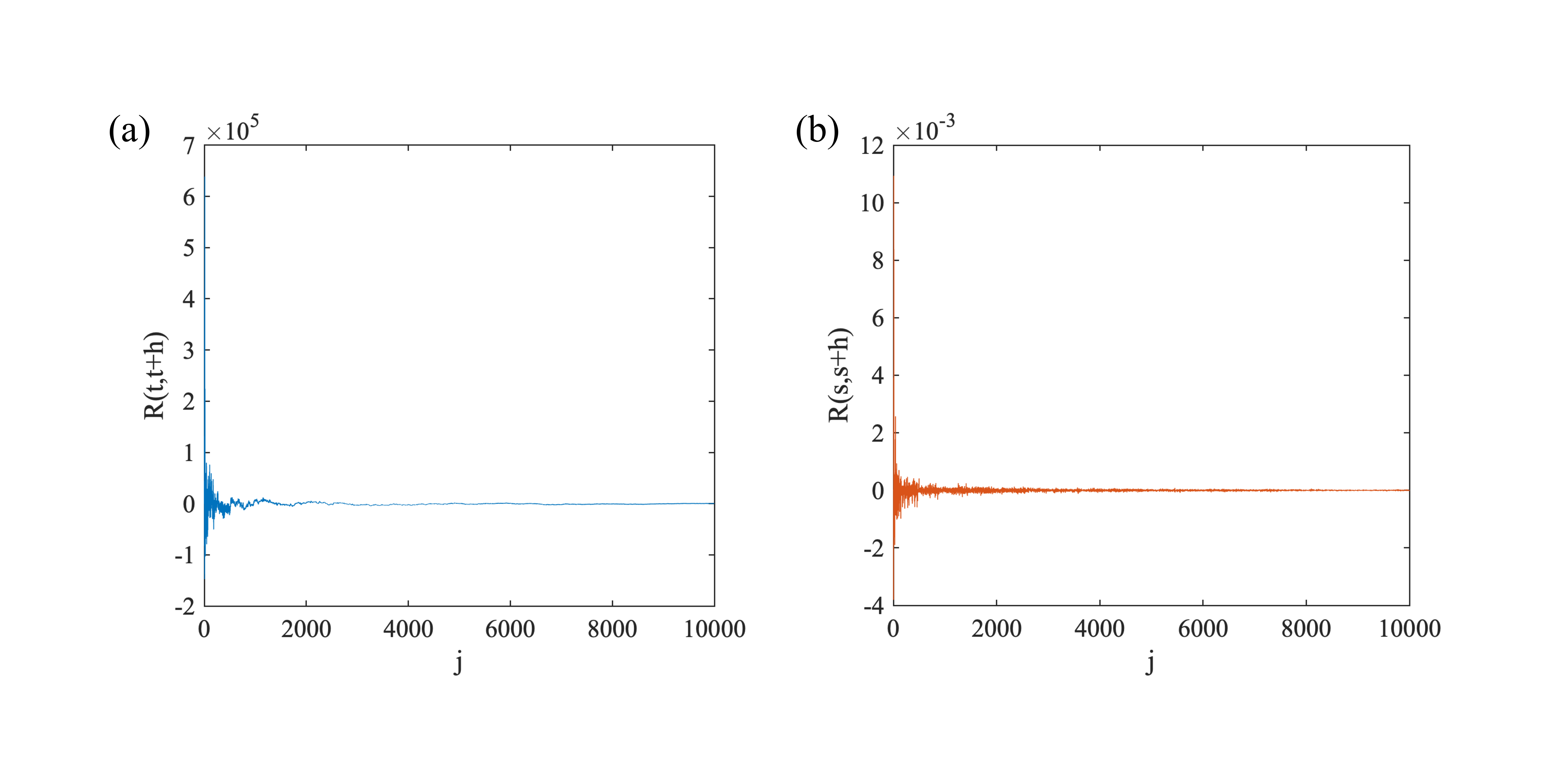}}
\caption{Plots of nonlinear response functions of two different types of FDRs. (a) Time-dependent dynamics of $R(t, t+h)$ calculated using Eq. (21). (b) Time-dependent dynamics of of $R(s, s+h)$ calculated using Eq. (29).} 
\end{figure}

\begin{figure}[htbp]
\centerline{\includegraphics[width=13cm]{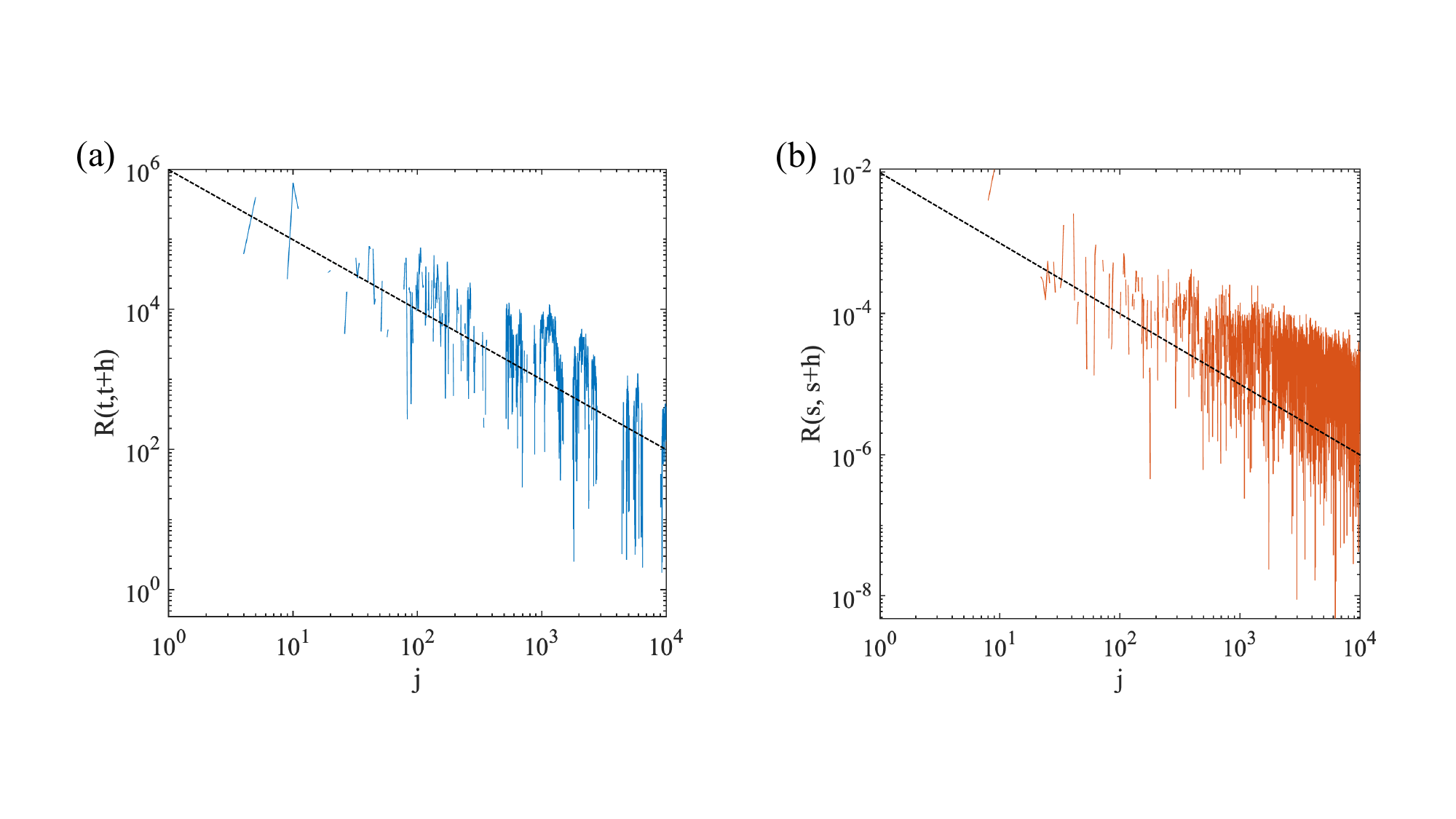}}
\caption{Exponential decay of nonlinear response functions. (a) Log-log plot of the time-dependent dynamics of $R(t, t+h)$ calculated using Eq. (21). (b) Log-log plot of the time-dependent dynamics of of $R(s, s+h)$ calculated using Eq. (29). The black dotted lines in each log-log plot represents the scaling of $t^{-1}$.}.
\end{figure}             

\section{Discussion}
We demonstrated that the conformal transformation governed by the LDE is applied for the decomposition of the operators appearing in the analyses of nonlinear GLE. This result is regarded as an alternative of the conventional Mori-Zwanzig method. The advantage of this approach is the computability of the operator $\mathcal{P}_n$, which enables the explicit numerical calculations of FDR. We should also notice that the present method and obtained results rely on the fact that the Brownian motion modeled by the GLE is appropriately measured by conformal geometry determined by LDE. Therefore, the comparison with the experimental data are required for further studies. However, because the present formalism satisfies the essential restrictions from the previous studies of the statistical-mechanical theory, we showed that the concepts from Loewner theory are useful for the analyses of the real-world settings of the diffusion process. This includes the reinterpretation of the Brownian motion in the colored noise environment. In the present formalism, by defining the Loewner entropy, the diffusion process with memory effects are deduced to that having the microcanonical ensemble. This theoretical fact supports the treatments of the FDRs in non-equilibrium states using the additive ensembles. This is enabled by the unique property of the Loewner maps and this type of conformal effect is not considered in the previous statistical-mechanical theory. In this sense, we showed that the reduction of the nonlinearity using the conformal map is enabled by Loewner map.    

\section{Conclusion}
In this paper, we have demonstrated the derivation of a nonlinear GLE using a modified Mori-Zwanzig method based on the conformal transformation governed by the LDE. We compared the different types of FDRs which were reformulated using Loewner theory to obtain the scaling laws of the energy dissipation of the diffusion process generated by GLE in non-equilibrium state. By applying a memory kernel corresponding to the cell migration experiment, we performed the numerical simulation to observe the time-dependent behavior of two types of the FDRs (Kubo-type, and a generalized type). As we showed, in this statistical-mechanical formalism, the concept of Loewner entropy $S_{Loew}$ plays a crucial role that is used for the reduction of the nonlinearity in the Brownian motion in the colored noise environment. Further studies are required for the comparison between the theoretical results and experimental results in physics, chemistry and biology.

\section*{Acknowledgments}
This work was performed when I am in charge of year-round lectures at College of Art, Nihon University. I would deeply acknowledge the academic staffs and students at the University, who have encouraged my academic activities.  

\section*{ORCID}

\noindent Yusuke K. Shibasaki - \url{https://orcid.org/0000-0001-9491-4118}

\appendix

\section{Dyson's operator identity method}
We here show the derivation of Eq. (11) from Eqs. (9) and (10), which is called Dyson's operator identity method. The method remarked below is essentially the same as that found in Ref \cite{vrugt2020}. Let us consider the quantity,  
\begin{equation}
A(t)=\exp(-t\mathcal{L})\exp(t\mathcal{Q}_n\mathcal{L}).
\end{equation}
The time derivative of $A(t)$ is expressed as: 
\begin{equation}
\frac{\partial}{\partial t} A(t)=-\exp(-t\mathcal{L}) \mathcal{P}_n \mathcal{L}\exp(t\mathcal{Q}_n\mathcal{L}).
\end{equation}
Integrating Eq. (A.2) with respect to $t$ using the initial condition $A(0)=1$, we obtain
\begin{equation}
A(t) = 1 -\int_0^t \exp(-t'\mathcal{L})\mathcal{P}_n \mathcal{L} \exp(t'\mathcal{Q}_n\mathcal{L}t')dt'. 
\end{equation}
Multiplying $\exp{\left(t\mathcal{L}\right)}$ to Eq. (A.3), we obtain the following 
\begin{equation}
\exp{\left(t\mathcal{Q}_n\mathcal{L}\right)}=\exp{\left(t\mathcal{L}\right)}+\int_0^t \exp[(t-t')\mathcal{L}]\mathcal{P}_n \mathcal{L}\exp(t\mathcal{Q}_n\mathcal{L}t')dt'
\end{equation}
Considering the time derivative of Eq. (A.4), we obtain Eq. (11) in the main text.

\section{Time coordinate change along the Loewner curve}
In this study, we use the time coordinate change from $t$ to $s$ as a technique to obtain the equilibrium ensemble of the observed variable. We here remark the basic procedure of this method, which is also used in refs \cite{shibasaki2022, shibasaki2024}. The techniques of the backward Loewner evolution shows the stochastic dynamics of the curve $\gamma_{[0,s]}$ generated by the Loewner equation in Eq. (1) is represented by those of $(X(s), Y(s))\in \mathbb{R}^2$ determined by the following differential equations: 
\begin{equation}
\begin{split}
\frac{d}{ds}X(s) &= - \frac{2X(s)}{X(s)^2+Y(s)^2} - \frac{d}{ds}U(s),\\
\frac{d}{ds}Y(s) &= \frac{2Y(s)}{X(s)^2+Y(s)^2}.
\end{split}
\end{equation}
Let us consider the substitution $Y(s)=t(s)$. Accordingly, we obtain the following:
\begin{equation}
\begin{split}
\frac{d}{ds}X(s) &= - \frac{2X(s)}{X(s)^2+t(s)^2} - \eta_s(t),\\
\frac{d}{ds}t(s) &= \frac{2t(s)}{X(s)^2+t(s)^2}.
\end{split}
\end{equation}
Here, we regard that $\frac{d}{ds}U(s)=\eta_s(t)$, which we referred to as the Loewner driving force. In Eqs. (22) and (26), we used the relation expressed by the second equation of Eq. (B.2). In addition, the substitution $X(s)=v_n$ corresponds the case of the analyses of the main text.

\end{document}